\documentclass[12pt]{article}
\usepackage{amsfonts}
\newcommand{\vf}{\varphi}
\newcommand{\be}{\begin{equation}}
\newcommand{\ee}{\end{equation}}
\newcommand{\ba}{\begin{eqnarray}}
\newcommand{\ea}{\end{eqnarray}}
\newcommand{\no}{\nonumber\\}
\newcommand{\bi}{\bibitem}
\newcommand{\vect}[1]{\stackrel{\rightarrow}{#1}}
\def\openone{\leavevmode\hbox{\small1\kern-3.8pt\normalsize1}}%

\usepackage{amsfonts}
\newcommand{\ab}[1]{{{\ensuremath{\mathbb{#1}}}}}
\begin{document}
\title{Non-existence  of  f-symbols in  generalized Taub-NUT 
spacetimes}
\author{Ion I. Cot\u{a}escu 
\thanks{E-mail:~~~ cota@physics.uvt.ro}\\
{\small \it West University of Timi\c soara,}\\
{\small \it V. P\^ arvan Ave. 4, RO-1900 Timi\c soara, Romania}
\and
Mihai Visinescu 
\thanks{E-mail:~~~ mvisin@theor1.theory.nipne.ro}\\
{\small \it Department of Theoretical Physics,}\\
{\small \it National Institute for Physics and Nuclear Engineering,}\\
{\small \it Magurele, P.O.Box MG-6, Bucharest, Romania}}
\date{\today}
%\date{ }
\maketitle

\begin{abstract}
In a previous article it was proved that the extensions of the Taub-NUT 
geometry do not admit Killing-Yano tensors, even if they possess 
St\"{a}ckel-Killing tensors. Here the analysis is taken further, and it 
is shown that, in general, this class of metrics does not even admit 
f-symbols. The only exception is the original Taub-NUT metric which 
possesses four Killing-Yano tensors of valence two.
\end{abstract}
\section{Extended  Taub-NUT spaces}
The Euclidean Taub-NUT metric is involved in many modern studies in 
physics \cite{Ma,AH}. From the view point of dynamical systems, the 
geodesic motion in Taub-NUT metric is known to admit a Kepler-type 
symmetry \cite{GM,GR,FH,CFH}. One can actually find the so called 
Runge-Lenz vector as a conserved vector in addition to the angular 
momentum vector. As a consequence, all the bounded trajectories are 
closed and the Poisson brackets among the conserved vectors give rise 
to the same Lie algebra as the Kepler problem has, depending on the 
energy. Thus the Taub-NUT metric provides a non-trivial generalization 
of the Kepler problem.

Iwai and Katayama \cite{IK} generalized the Taub-NUT metric so that it 
still admit a Kepler-type symmetry. The extended Taub-NUT metric, denoted 
by $ds^2_K$, is defined on {\ab R}$^4 - \{0\}$ by
\be\label{sg}
ds^2_K=f(r)(dr^2+r^2d\theta^2+r^2\sin^2\theta\, d\vf^2)
+g(r)(d\chi+\cos\theta\, d\vf)^2 
\ee
where $r>0$ is the radial coordinate, 
the angle variables $(\theta,\vf,\chi)$ parametrize the unit sphere
$S^3$ with $ 0\leq\theta<\pi, 0\leq\vf<2\pi, 
0\leq\chi<4\pi$, and $f(r)$ and $g(r)$ are  functions given, with 
constants $a,b,c,d$, by
\be\label{fg}
f(r) = { a + b r\over r}  ~~,~~ g(r) = { a r + b r^2 \over 
1 + c r + d r^2}\,.
\ee
If one takes the constants $c={2 b\over a}, d = {b\over a}^2$ with
$4m = {a\over b}$, the extended Taub-NUT metric becomes the original
Euclidean Taub-NUT metric up to a constant factor.

Spaces with a metric of the above form  have an isometry group 
$ SU(2)\times U(1)$. The four Killing vectors are \cite{GM}:
\be \label{da}
D_A=R_A^\mu\,\partial_\mu,~~~~A=0,1,2,3,~~~~ \mu=(r,\theta,\vf,\chi) 
\ee
with
\ba \label{ra}
R_0&=& (0,0,0,1),\no
R_1&=& (0,-\sin\vf,-\cot\theta\cos\vf,\csc\theta\cos\vf),\no
R_2&=& (0,\cos\vf,-\cot\theta\sin\vf,\csc\theta\sin\vf),\no
R_3&=& (0,0,1,0)\,.
\ea

$D_0$ which generates the $U(1)$ of $\chi$ translations, commutes 
with the other Killing vectors. The remaining three Killing vectors 
$D_i,~~ i=1,2,3$, corresponding to
the invariance of the metric (\ref{sg}) under spatial
rotations.

The conserved quantities along geodesics are homogeneous functions in 
momentum $p_\mu$ which commute with the Hamiltonian
\be\label{h}
H = {1\over 2} g^{\mu\nu} p_\mu p_\nu
\ee
in the sense of Poisson brackets.

The usual constants of motion for particles moving in this background 
are linear in the four momentum $p_\mu$:
\be \label{ja}
J_A = R^\mu_A ~ p_\mu\,.
\ee

The conserved quantity corresponding to the cyclic variable $\chi$ is 
given by
\be \label{q}
q = g(r) (\dot\theta + \cos\theta \dot\vf)
\ee
where the over-dot denotes an ordinary proper-time derivative. For 
negative mass models, this quantity can be interpreted as the 
``relative electric charge" \cite{AH,GM,GR,FH,CFH}. For $q \in${\ab R}
fixed, the geodesic flow system for the Taub-NUT metric is reduced to  
a Hamiltonian system on $T^*(${\ab R}$^3 - \{0\}) \cong
(${\ab R}$^3 - \{0\})\times ${\ab R}$^3$ with Cartesian coordinates 
$(x_j,p_j),\, j=1,2,3$ and the Hamiltonian function
\be\label{en}
H_q = {\vect{p}^{~2}\over 2 f(r)} + {q^2 \over 2 g(r)} \,.
\ee

Owing to the obvious spherical symmetry, the angular momentum vector 
\be \label{j}
\vect{J}=\vect{x}\times\vect{p}\,+\,q\,{\vect{x}\over r} 
\ee
is a conserved vector. 

The remarkable result of Iwai and Katayama is that the extended Taub-NUT 
space (\ref{sg}) still admits a conserved vector, quadratic in 
$4$-velocities, analogous to the Runge-Lenz vector of the following form
\be\label{rl}
\vect{S} = \vect{p} \times \vect{J} + \kappa {\vect{x} \over r}\,.
\ee
Setting the value of the Hamiltonian $H_q$ to $E$ the constant $\kappa$ 
involved in the Runge-Lenz vector (\ref{rl}) is
\be
\kappa = - a\,E + {1\over 2} c\,q^2.
\ee

The Poisson brackets between the components of $\vect{J}$ and 
$\vect{S}$ are similar to the relations known for the original 
Taub-NUT metric \cite{IK}. In particular
\be\label{jj} 
\{J_i, S_j\} = \epsilon_{ijk} S_k \,.
\ee

\section{Killing-Yano tensors and f-symbols}
The explicit form of the $3$-vector $\vect{S}$ (\ref{rl}) is
\be\label{rlvec}
\vect{S} = \left[\left(f^2 - {a f\over 2 r}\right)\dot{\vect{x}}^2
+\left({c\over 2 r}-{a\over 2gr}\right) q^2\right]\vect{x} - {q\over r} 
\vect{x}\times\vect{p} - f^2 r\dot{r}\dot{\vect{x}}.
\ee
It is straightforward to verify that the components of the $3$-vector 
$\vect{S}$ are St\" ackel-Killing tensors of valence $2$ \cite{MV}:
\be\label{sk}
S_{i(\mu\nu;\lambda)} = 0~~,~~~i=1,2,3
\ee
confirming the expectations. Actually such integrals of motion are 
related to hidden symmetries of the manifold, which manifest 
themselves as St\" ackel-Killing tensors

There are geometries where 
the St\" ackel-Killing tensor can have a certain root represented by 
Killing-Yano tensors. We recall that a tensor $f_{\mu\nu}$ is a 
Killing-Yano tensor of valence $2$ if it is totally antisymmetric and 
satisfies the equation \cite{Ya}
\be\label{ky}
f_{\mu(\nu;\lambda)} = 0\,.
\ee

In the original Taub-NUT geometry there are four Killing-Yano tensors 
\cite{GR}. Three of these are covariantly constant
\ba\label{fi}
f_i &=&8m(d\chi + \cos\theta d\varphi)\wedge dx_i -
\epsilon_{ijk}\left(1+\frac{4m}{r}\right)\, dx_j \wedge dx_k,\no
D_\mu f^\nu_{i\lambda} &=&0~, ~~~~i,j,k=1,2,3\,.
\ea
They are mutually anticommuting and square the minus unity.
Thus they are complex structures realizing the quaternion algebra and 
the original Taub-NUT manifold is hyper-K\" ahler. 

In addition to the above vector-like Killing-Yano tensors there is also 
a scalar one
\be\label{fy}
f_Y =8m(d\chi + \cos\theta  d\varphi)\wedge dr +
4r(r+2m)\left(1+\frac{r}{4m}\right)\,\sin\theta  d\theta \wedge d\varphi
\ee
which is not covariantly constant.

In the original Taub-NUT space the components $S_{i\mu\nu}$ involved 
with the Runge-Lenz type vector (\ref{rl}) can be expressed as 
symmetrized products of the Killing-Yano tensors $f_i$ (\ref{fi}) and
$f_Y$ (\ref{fy}) \cite{VV}: 
\be\label{kff}
S_{i\mu\nu} - \frac{1}{8m} (R_{0\mu} R_{i\nu} + R_{0\nu} R_{i\mu}) = 
m\left( f_{Y\mu\lambda}
{{f_{i}}^\lambda}_\nu + f_{Y\nu\lambda} {{f_{i}}^\lambda}_\mu
\right)\,. 
\ee

In fact, only the product of Killing-Yano tensors $f_i$ and $f_Y$ leads 
to non-trivial St\" ackel-Killing tensors, the last term in the left 
hand side of (\ref{kff}) being a simple product of Killing vectors. 
This term is usually added to write the Runge-Lenz vector in the 
standard form (\ref{rl}).
The existence of the Killing-Yano tensors is connected with the 
appearance  of additional supersymmetries in the usual $N=1$ 
supersymmetric extension of point particle mechanics in curved 
spacetime.

However, in general, the St\" ackel-Killing tensors involved in the 
Runge-Lenz vector cannot be expressed as a product of Killing-Yano 
tensors \cite{MV}. The extensions of the Taub-NUT geometry do not admit 
a Killing-Yano tensor, even if they possess St\" ackel-Killing tensors. 
Therefore the only exception is the original Taub-NUT metric.

If the Killing-Yano tensors are missing, to take up the question of the 
existence of extra supersymmetries and the relation with the constants 
of motion it is necessary to enlarge the approach to Killing equations 
(\ref{sk}) and (\ref{ky}). In \cite{GRH} it was shown that it is 
possible to make a weaker demand that an extra supersymmetry exists. It 
was shown that supersymmetries depend on the existence of a second-rank 
field $f_{\mu\nu}$ called {\it f-symbol}. The f-symbols satisfy 
relation (\ref{ky}), but the covariant tensor fields $f_{\mu\nu}$ need 
not necessarily be antisymmetric.

The symmetric part of an f-symbol is the tensor
\be\label{sym}
S_{\mu\nu} = {1\over 2} ( f_{\mu\nu} + f_{\nu\mu})
\ee
which is a St\" ackel-Killing tensor satisfying  Killing 
equation (\ref{sk}). The antisymmetric part
\be\label{as}
A_{\mu\nu} = {1\over 2} ( f_{\mu\nu} - f_{\nu\mu})
\ee 
obeys the condition
\be\label{eqa}
D_\mu A_{\nu\lambda} + D_\nu A_{\mu\lambda} = D_\lambda S_{\mu\nu}\,.
\ee

\section{Non-existence of f-symbols}
Bearing in mind that any extension of the original Taub-NUT space does 
not admit Killing-Yano tensors, in what follows we are in search of 
f-symbols. For this purpose we seek for solutions of equation 
(\ref{eqa}) with appropriate St\" ackel-Killing tensors in the right 
hand side.

Taking into account that the Runge-Lenz vector $\vect{S}$ (\ref{rl}) 
transforms as a vector under rotations generated by $\vect{J}$ 
according to (\ref{jj}), and, in view of the decomposition (\ref{kff}) 
for the original Taub-NUT space, we are looking for vector and scalar 
f-symbols. Consequently, in the right hand side of equation (\ref{eqa}) 
we must have St\" ackel-Killing tensors with the corresponding behavior 
under $3$-dimensional rotations.

First we consider equations (\ref{eqa}) for a scalar f-symbol with 
scalar St\" ackel-Killing tensors in the right hand side. From the 
Killing vectors (\ref{da}) we can form a trivial scalar St\" 
ackel-Killing tensor
\be\label{scal}
\alpha (R_{0\mu}R_{0\nu}) + \beta (R_{1\mu}R_{1\nu} + R_{2\mu}R_{2\nu} 
+ R_{3\mu}R_{3\nu})
\ee
with $\alpha$ and $\beta$ constants.

{}From the set of partial differential equations (\ref{eqa}) we shall 
select the following ones:
\ba
\mu = \nu = r ~,~ \lambda = \theta :~~~~~~~
&&{\partial A_{r\theta}\over \partial r} - \left({1\over r} + 
{f'\over f}\right) 
A_{r\theta} = 0\, ,\label{1}\\
\mu = \nu = \theta ~,~ \lambda = r :~~~~~~~
&&{\partial A_{r\theta}\over \partial \theta} = - \beta\left(f^2 r^3 + 
{f f'\over 2} r^4\right)\, ,\label{4}\\
\mu = \nu = r ~,~ \lambda = \chi :~~~~~~~
&&{\partial A_{r\chi}\over \partial r} - \left({f'\over 2 f} + 
{g'\over 2 g}\right) 
A_{r\chi} = 0\, ,\label{3}\\
\mu = \nu = \chi ~,~ \lambda = r :~~~~~~~
&&{\partial A_{r\chi}\over \partial \chi} = - {g g'\over 2} (\alpha + 
\beta)\label{12}
\ea
where a prime denotes a derivative with respect to $r$.

The integrability condition for the pair of equations (\ref{3}) and 
(\ref{12}) is
\be\label{312}
(\alpha + \beta) \left[({g'}^2 + g g'') - g g' \left({f'\over 2 f} + 
{g'\over 2 g}\right)\right] = 0\,.
\ee

But for the functions $f(r)$ and $g(r)$ given by (\ref{fg}) which 
ensure the Kepler-type symmetry of the extended Taub-NUT space, the 
above relation can be satisfied only for
\be\label{albe}
\alpha + \beta = 0\,.
\ee

Now from the integrability condition for the pair of equations (\ref{1}) 
and (\ref{4}) we get 
\be\label{be}
\beta\left(f^2 r^3 + 
{f f'\over 2} r^4\right)'
- \beta \left({1\over r} + {f'\over f}\right) \left(f^2 r^3 + 
{f f'\over 2} r^4\right) = 0
\ee
which, again, for the form (\ref{fg}) of the function $f(r)$ leads to
\be\label{beta}
\beta = 0\,.
\ee

Therefore the right hand side (\ref{scal}) of  equation (\ref{eqa}) 
vanishes in the scalar case. But without a symmetric part, the 
equations for f-symbols lead precisely to equations (\ref{ky}) for 
Killing-Yano tensors. Now, in conjunction with the absence 
of the Killing-Yano tensors on extended Taub-NUT space (with the 
exception of the original Taub-NUT metric) we conclude that not even 
scalar f-symbols exist.

For a vectorial f-symbol we must consider for the right hand side of 
equation (\ref{eqa}) a combination between the component $S_{i\mu\nu}$ 
of the Runge-Lenz vector (\ref{rl}) and the trivial St\" ackel-Killing 
tensor of the form $ R_{0\mu}R_{i\nu} + R_{0\nu}R_{i\mu}$ with $i= 
1,2,3$. Again a detailed analysis of the integrability conditions for 
equations (\ref{eqa}), similar to that done above, leads to the 
conclusion that vectorial f-symbols do not exist.
\section{Conclusions}
In supersymmetric quantum mechanics models with standard supersymmetry, 
the supercharges $Q_a$ close on Hamiltonian $H$ so that we have
$\{Q_a,Q_b\} = 2 \delta_{ab}H, ~~~ a,b = 1,...N.$

In some cases one can find additional hidden supercharges of the 
nonstandard form \cite{GRH,Mac} involving the structure constants of a 
Lie algebra and perhaps Killing-Yano tensors. The appearance of the 
Killing-Yano tensors is not surprising since they play a role in the 
existence of hidden symmetries \cite{GR,Ta}.
The analysis of the f-symbols presented in \cite{GRH} shows that the 
Killing-Yano and St\" ackel-Killing tensors belong to a larger class of 
possible structures which generate generalized supersymmetry algebras.
Unfortunately, to our knowledge, there is no explicit example of 
f-symbols in the literature.

The absence of Killing-Yano tensors or eventually f-symbols in extended 
Taub-NUT geometry, in spite of the existence of hidden symmetries in 
this class of spaces, is quite troublesome. For example in the formalism 
of pseudo-classical spinning point particles using anticommuting 
Grassmann variables to describe the spin degrees of freedom 
\cite{GRH,BM}, the Killing-Yano tensors or f-symbols play an essential 
role in the study of generalized Killing equations. The construction of 
the constants of motion for spinning particles is severely hampered by 
the absence of the f-symbols. Moreover, in the absence of these 
geometrical objects,  it is not clear how to compute the spin 
corrections to the conserved quantities corresponding to hidden 
symmetries or even if these corrections exist.

\end{document}